\documentclass[a4paper]{article}
\usepackage{authblk}

\usepackage[utf8]{inputenc}
\usepackage{graphicx}

\usepackage[mediumspace,mediumqspace,squaren]{SIunits}

\begin{document}

\title{Comparative study of RBE and cell
survival fractions for $^{1}$H, $^{4}$He, $^{12}$C and $^{16}$O beams using Geant4 and Microdosimetric Kinetic model}

\author[1,2]{Lucas~N~Burigo%
  \thanks{Electronic address: \texttt{burigo@fias.uni-frankfurt.de}; Corresponding author}}
\author[1,3]{Igor~A~Pshenichnov}
\author[1,4]{Igor~N~Mishustin}
\author[1,2]{Marcus~Bleicher}

\affil[1]{Frankfurt Institute for Advanced Studies, Johann Wolfgang Goethe University, 60438 Frankfurt am Main, Germany}
\affil[2]{Institut f\"ur Theoretische Physik, Johann Wolfgang Goethe University, 60438 Frankfurt am Main, Germany}
\affil[3]{Institute for Nuclear Research, Russian Academy of Sciences, 117312 Moscow, Russia}
\affil[4]{Kurchatov Institute, Russian Research Center, 123182 Moscow, Russia}

\maketitle

\begin{abstract}
Beams of $^{4}$He and $^{16}$O nuclei are considered for ion-beam cancer therapy as alternative options to protons and $^{12}$C nuclei. Spread-out Bragg peak (SOBP) distributions of physical dose and relative biological effectiveness for 10\% 
survival are calculated by means of our Geant4-based Monte Carlo model for Heavy Ion Therapy (MCHIT) and the
modified microdosimetric kinetic model. The depth distributions of cell survival fractions are calculated for 
$^{1}$H, $^{4}$He, $^{12}$C and $^{16}$O for tissues with normal (HSG cells), low and high radiosensitivity.
In each case the cell survival fractions were compared separately for the target volume, behind and in front of it.
In the case of normal radiosensitivity $^{4}$He and $^{12}$C better spare tissues 
in the entrance channel compared to protons and $^{16}$O. The cell survival fractions calculated, respectively, for the entrance channel and target volume are similar for $^{4}$He and $^{12}$C.  When it is important to spare 
healthy tissues located after the distal edge of the SOBP plateau, $^{4}$He can be recommended due to reduced nuclear fragmentation of these projectiles. No definite advantages of $^{16}$O with respect to $^{12}$C were found, 
with the except of an enhanced impact of these heavier projectiles on radioresistant tumors. 
\end{abstract}

\maketitle

\section{Introduction}

The advantage of charged particles, in particular, protons and
carbon nuclei, used for radiation therapy of cancer consists in elevated dose delivered 
at the end of projectile range in tissues. The plateau of the depth-dose distribution at the 
entrance of a monoenergetic beam terminates with a sharp Bragg peak which can be targeted at 
the tumor. Such a dose profile helps to spare healthy tissues located in front of the 
tumor as well as beyond the projectile range. Since a set of beam energies is typically used in 
treatments to cover the whole tumor volume, the resulting dose 
distribution is characterized by a spread-out Bragg peak (SOBP)~\cite{Kase2006a,Gueulette2007,Schardt2010} 
with a wide domain of elevated dose. 

The damage of healthy tissues during therapy can be essentially reduced if the ratio between the relative biological effectiveness (RBE) values 
at the SOBP region and plateau is favorable.  As recognized almost 40 years ago in radiobiological experiments 
with SOBP beams of light nuclei performed at Berkeley~\cite{Chapman1977}, this RBE ratio is greater 
than 1 and increases with ion charge up to carbon. It was also found that this ratio decreases 
for Ne and becomes less than 1. In 1994 first patient treatments with beams of carbon nuclei started in Japan at 
National Institute of Radiological Sciences (NIRS)~\cite{Kanai1999} and in 1997 in Germany at 
Gesellschaft für Schwerionenforschung (GSI)~\cite{Schardt2010}. Later   
the advantages of $^{12}$C with respect to $^{3}$He and $^{20}$Ne were confirmed in experiments at NIRS with these nuclear beams~\cite{Furusawa2000}. As explained~\cite{Matsufuji2007}, the RBE of \unit{135}{\mega\electronvolt\per\atomicmass} $^{12}$C beam with the linear energy transfer (LET)  
of \unit{65}{\kilo\electronvolt\per\micro\meter} was found similar to the RBE of neutrons which have been used 
for treatment at NIRS during last 20 years. This similarity also motivated the choice of carbon nuclei for treatments at NIRS.  In the last decades localized tumors have been 
successfully treated with beams of carbon nuclei at several facilities constructed in 
Japan and Germany~\cite{Tsujii2004,Schulz-Ertner2007,Durante2010,Combs2010a,Jensen2011,Kamada2012}.  

Despite of the broad clinical experience collected worldwide with proton and carbon-ion beams,
other light nuclei can be also considered as future therapy options. The radiobiological properties of proton 
and $^{12}$C beams were compared in several studies, see e.g.~\cite{Wilkens2008,Suit2010}. However, 
less attention has been paid so far to $^{4}$He or $^{16}$O and to their comparison 
with protons and $^{12}$C. 
There exist several clinical rationale behind the use of $^4$He and $^{16}$O for therapy:
\begin{itemize}
 \item[i)] $^4$He and $^{16}$O beams have a reduced lateral spread of the dose distribution compared to protons;
 \item[ii)] their RBE in the target volume is higher compared to protons;  
 \item[iii)] lower dose in the tail region and lower RBE in the plateau is expected for $^4$He compared to $^{12}$C due to reduced nuclear fragmentation of $^4$He; 
 \item[iv)] $^{16}$O is a promising option for hypoxic tumors as it provides a higher dose averaged LET in the target volume compared to $^{12}$C.
\end{itemize}

The choice of ion species and their energies at each new particle therapy facility~\cite{Brahme2001,Svensson2004,Lundkvist2005} essentially depends on LET and RBE of the  
projectiles under consideration. The distributions of dose, LET and dose 
averaged LET of $^{1}$H, $^{4}$He, $^{6,7}$Li, $^{8}$Be, $^{10}$B, $^{12}$C, 
$^{14}$N and $^{16}$O nuclei of therapeutic energies were studied~\cite{Kempe2007,Kantemiris2011}
by means of Monte Carlo simulations with SHIELD-HIT and FLUKA codes, respectively.
Similar distributions for $^{1}$H, $^{3}$He, $^{12}$C, $^{20}$Ne and $^{58}$Ni were 
calculated~\cite{Pshenichnov2008} with the Geant4 toolkit to
study their dependence on the ions mass, charge and energy.  
It was assumed~\cite{Kempe2007} that ions with LET above  
\unit{20}{\kilo\electronvolt\per\micro\meter} should be used for efficient 
cancer therapy as such species induce on average two and more 
double strand breaks in DNA close to each other. However, as pointed out in the same work, this limit is 
not sharp and ought to vary with ion mass and charge.  This indicates that the RBE of respective ions has 
to be additionally considered for an accurate comparison of their biological action. Indeed, as shown in our recent
work~\cite{Burigo2014}, there is no direct correspondence between RBE and the frequency-mean linear 
energy $\bar{y}_f$, which represents LET, for monoenergetic beams of therapeutic energies.  
Similar RBE values were estimated~\cite{Burigo2014} at the peak and plateau regions which are 
characterized, however, by very different $\bar{y}_f$. 

As demonstrated~\cite{Burigo2014}, microdosimetry spectra for monoenergetic $^{1}$H, $^{4}$He, $^{7}$Li and $^{12}$C
nuclei propagating in a water phantom can be accurately described with our 
Monte Carlo model for Heavy-Ion Therapy (MCHIT)~\cite{Pshenichnov2010}, and this model
coupled with the Microdosimetric Kinetic (MK) model~\cite{Hawkins2003,Kase2006} can be used to calculate 
the respective RBE profiles. In the present work we evaluate $^{4}$He and $^{16}$O for cancer therapy as complementary options to $^{1}$H and $^{12}$C  by considering the biological dose distribution with a \unit{6}{\centi\metre} SOBP delivered by these four projectiles. This is an important prerequisite for planning radiobiological experiments with $^{4}$He and $^{16}$O SOBP beams and extending existing treatment planning systems to operation with $^{4}$He and $^{16}$O.

The experimental data~\cite{Matsufuji2007} collected at HIMAC for the \unit{6}{\centi\metre} SOBP obtained by the moderation of a \unit{290}{\mega\electronvolt\per\atomicmass} $^{12}$C beam are used as a reference. 
The results of microdosimetry simulations are validated by comparison with microdosimetry data collected
for $^{1}$H, $^{4}$He and $^{12}$C beams~\cite{Kase2006}. This makes possible to predict the RBE and cell survival profiles for $^{16}$O beams and compare all four ion species in a common framework.

\section{Materials and methods}

\subsection{Monte Carlo modeling of propagation of ions in water}

With the Monte Carlo model for Heavy Ion Therapy (MCHIT)~\cite{Pshenichnov2005,Pshenichnov2010} the 
propagation of accelerated protons and light nuclei of therapeutic energies in tissue-like media can be 
simulated. The model is based on the Geant4 toolkit~\cite{Agostinelli2003,Allison2006} and takes into account all major physics processes relevant to the interactions of beam particles with these materials. 
The Geant4 version 9.5 with patch 02 is used to build the present version of MCHIT, which also 
simulates the interactions of various particles with walled and wall-less Tissue Equivalent Proportional Counters (TEPC) thus providing respective microdosimetry distributions. 

The ionization of atoms and multiple Coulomb scattering on nuclei of the media are the most important electromagnetic processes to simulate the energy loss and straggling of primary and secondary charged particles. Two predefined physics lists for electromagnetic processes are employed in MCHIT, namely, G4EmStd (which uses the ``Standard Electromagnetic Physics Option 3'') and G4EmPen (which uses the Penelope models for low-energy processes). The low-energy thresholds for production of $\delta$-electrons are 990~eV for G4EmStd and 100~eV for G4EmPen. However, in order to reduce the CPU time, cuts in range and energy 
thresholds for particle production are set differently for water, plastic of the TEPC wall and the TEPC sensitive volume filled with tissue-equivalent gas~\cite{Burigo2013}.  A customized physics list, G4EmPen+IonGas, which is based on G4EmPen and the models describing the ionization of gas media by ions, can be also used in calculations. As demonstrated~\cite{Burigo2014}, G4EmStd and G4EmPen provide statistically equivalent results for microdosimetry spectra, which agree well with the distributions measured 
for $^4$He~\cite{Tsuda2012}, 
excluding the domains of low linear energy,  $y <$\unit{1}{\kilo\electronvolt\per\micro\metre}, and
around the maximum at $y\sim$\unit{15}{\kilo\electronvolt\per\micro\metre}. However, the agreement between the data and calculations is improved when G4EmPen+IonGas is used. Therefore, G4EmPen+IonGas is involved in MCHIT also in the present work to simulate electromagnetic processes inside the TEPC volume.

A therapeutic nuclear beam is attenuated in tissues due to the loss of beam nuclei in
nuclear fragmentation reactions, which generate secondary projectile and target fragments~\cite{Pshenichnov2010}.  Nuclear reactions are taken into account in MCHIT to reproduce this effect. As shown~\cite{Pshenichnov2010}, the build-up of secondary fragments produced by 
\unit{200}{\mega\electronvolt\per\atomicmass} and \unit{400}{\mega\electronvolt\per\atomicmass} $^{12}$C beams is generally well described with a customized physics list based on the Light Ion Binary Cascade model (G4BIC)~\cite{Folger2004} coupled with the Fermi break-up model (G4FermiBreakUp)~\cite{Bondorf1995} 
responsible for subsequent decays of excited nuclear fragments created at the first fast 
stage of nucleus-nucleus collisions. In the present work G4BIC is used for proton, helium 
and lithium beams, while the Quantum Molecular Dynamics model (G4QMD)~\cite{Koi2010} is involved in simulations with carbon and oxygen beams. More details on the physics processes and respective 
Geant4 models involved in modeling with MCHIT are given in our recent  publications~\cite{Pshenichnov2010,Burigo2013,Burigo2014}.

\subsection{Microdosimetry simulations and calculations of RBE for monoenergetic beams}
\label{Sec:monoenergetic}

The design and materials of specific TEPC models were thoroughly introduced in MCHIT. 
This made possible to simulate the microdosimetry
spectra measured with a walled TEPC at several positions inside a water phantom irradiated by \unit{185}{\mega\electronvolt\per\atomicmass} $^{7}$Li and \unit{300}{\mega\electronvolt\per\atomicmass} $^{12}$C beams~\cite{Martino2010} and study the impact of nuclear fragmentation reactions on these spectra~\cite{Burigo2013,Burigo2014}. 
The influence of the positioning of the TEPC with respect to the beam axis and the 
distortion of the spectra due to the pile-up of individual events were investigated.  
After correcting for such effects the calculated microdosimetry spectra agree well with the experimental data in general. 

Following the validation of MCHIT for microdosimetry of monoenergetic $^{7}$Li and $^{12}$C 
beams~\cite{Burigo2013,Burigo2014}, this model is applied to microdosimetry of SOBP dose distributions considered in the present work. 
In the measurements performed at HIMAC with a \unit{6}{\centi\metre} SOBP for 
\unit{160}{\mega\electronvolt} $^{1}$H, \unit{150}{\mega\electronvolt\per\atomicmass} $^{4}$He and \unit{290}{\mega\electronvolt\per\atomicmass} $^{12}$C~\cite{Kase2006} the data for frequency-mean lineal energy, $\bar{y}_f$, dose-mean lineal energy, $\bar{y}_d$, and saturation-corrected dose-mean lineal energy, $y^*$~\cite{ICRU1983} were collected. 
The microdosimetry spectra were measured with a walled TEPC corresponding to a tissue-equivalent sphere of 
\unit{1}{\micro\metre} in diameter. 

According to the linear-quadratic (LQ) model the fraction of cells $S$ survived  after the impact of  
the radiation dose $D$ is calculated as
\begin{equation}
S = \exp{\left[ -\alpha D -\beta D^2 \right]} \ .
\label{eq:SLQ}
\end{equation}
Following the modified MK model~\cite{Kase2006} applied to human salivary gland (HSG) tumor cells 
the parameter $\alpha$ is estimated as
\begin{equation}
\alpha = \alpha_0 + \frac{\beta}{\rho \pi r_d^2} y^* \ ,
\label{eq:alpha}
\end{equation}
with the constant term $\alpha_0$ = 0.13~Gy$^{-1}$ representing the initial slope of the survival fraction curve in the limit of zero LET and $\beta$ = 0.05~Gy$^{-2}$, $\rho = $1~g/cm$^3$ as the density of tissue and $r_d = $0.42~$\mu$m as the radius of a sub-cellular domain in the MK model. 
The dependence of the $\alpha$-parameter of the linear-quadratic model on $y^*$ rather than on $\bar{y}_d$
reflects the reduction of the RBE known as the saturation effect. It means that an excessive local energy deposition does not boost biological effects induced by high-LET particles~\cite{ICRU1983}.
As demonstrated~\cite{Kase2006}, the same value of parameter $\beta$ = 0.05~Gy$^{-2}$ 
can be used to fit the data on $S$ with Eq.~(\ref{eq:SLQ}) for X-rays and ions. This justifies 
the assumption of the MK model that $\beta$ is independent of LET. 

According to the LQ model the RBE$_{10}$ for 10\% survival of HSG cells is calculated using the following relation~\cite{Kase2011}:
\begin{equation}
{\rm RBE}_{10} = \frac{D_{10,R}}{D_{10}} = \frac{2\beta D_{10,R}}{\sqrt{\alpha^2-4\beta\ln{\left(0.1\right)}}-\alpha},
\label{eq:RBE}
\end{equation}
where $D_{10}$ is the 10\% survival dose of ions and $D_{10,R}$ = 5.0~Gy is the 10\% 
survival dose of the reference radiation (200~kVp X-rays) for HSG cells~\cite{Kase2011}. Finally,
the biological dose $D_{bio}$ is calculated from RBE$_{10}$ and physical dose:
\begin{equation}
D_{bio} = {\rm RBE}_{10}\  D.
\label{eq:biodose}
\end{equation}

\subsection{Composing SOBP profiles from a library of pristine Bragg peaks}\label{Sec:SOBPmethod}

The computing time required for treatment planing in carbon-ion therapy can be reduced by  
using pre-computed libraries of dose and RBE distributions for monoenergetic 
beams~\cite{Kramer2000,Kramer2000a,Jakel2001}.
The aim of the treatment planning is to find an optimum superposition of many
beams with their individual energy, position and intensity in order to 
obtain the prescribed biological SOBP dose profile. It is expected that a similar approach 
will be also suitable for other therapeutic beams, like $^{4}$He and $^{16}$O. 
Therefore, we implemented a common algorithm to calculate the relative weights of pre-defined 
monoenergetic beams of $^{1}$H, $^{4}$He, $^{12}$C and $^{16}$O to obtain flat biological 
SOBP distributions for each projectile as a product of the physical dose and RBE
calculated for mixed radiation field.

A library of depth-dose profiles and the corresponding microdosimetry spectra for different beam energies and nuclei were calculated by Monte Carlo simulations with MCHIT. They are used as input 
data for a procedure similar to one implemented at NIRS~\cite{Matsufuji2007}.
According to this procedure based on the theory of dual radiation action~\cite{Zaider1980} the survival fraction of cells exposed to mixed radiation is calculated as:
\begin{equation}
 S_{mix}(D) = \exp\left( -\alpha_{mix}D - \beta_{mix}D^2\right);
\end{equation}
\begin{equation}
 \alpha_{mix} = \sum f_i\alpha_{i}; \\
\end{equation}
\begin{equation}
 \sqrt{\beta_{mix}} = \sum f_i \sqrt{\beta_{i}}.
\end{equation}
Here $f_i$ is the weight coefficient (fraction) of the local physical dose of the $i$th monoenergetic beam 
which contribute to the total physical dose $D$, while $\alpha_i$ and $\beta_i$ are the parameters of the LQ model specific to $i$th monoenergetic beam. 
The parameters $\alpha_i$ and $\beta_i$ are calculated along the beam axis using MCHIT 
coupled with the modified MK model as described in Sec.~\ref{Sec:monoenergetic}. The resulting RBE$_{10,mix}$ for the mixed radiation is calculated from the survival fraction of cells 
$S_{mix}(D)$ and it also depends on the depth. RBE$_{10}$ and RBE$_{10,mix}$ 
for the monoenergetic and SOBP (mixed) beams, respectively, are considered in the following. 
They account for the relative biological effectiveness corresponding to 10\% survival of cells. 

A dedicated algorithm to obtain $f_i$ for a given biological SOBP was developed. It starts with the determination of the weight at the distal edge of the SOBP distribution and then 
calculates weights for less energetic beams by adjusting their contribution to provide a flat 
SOBP plateau.

\section{Results and discussion}

\subsection{Pristine Bragg peaks}\label{Sec:pristine}

\begin{figure}[htb!]
\centering
\includegraphics[width=0.65\columnwidth]{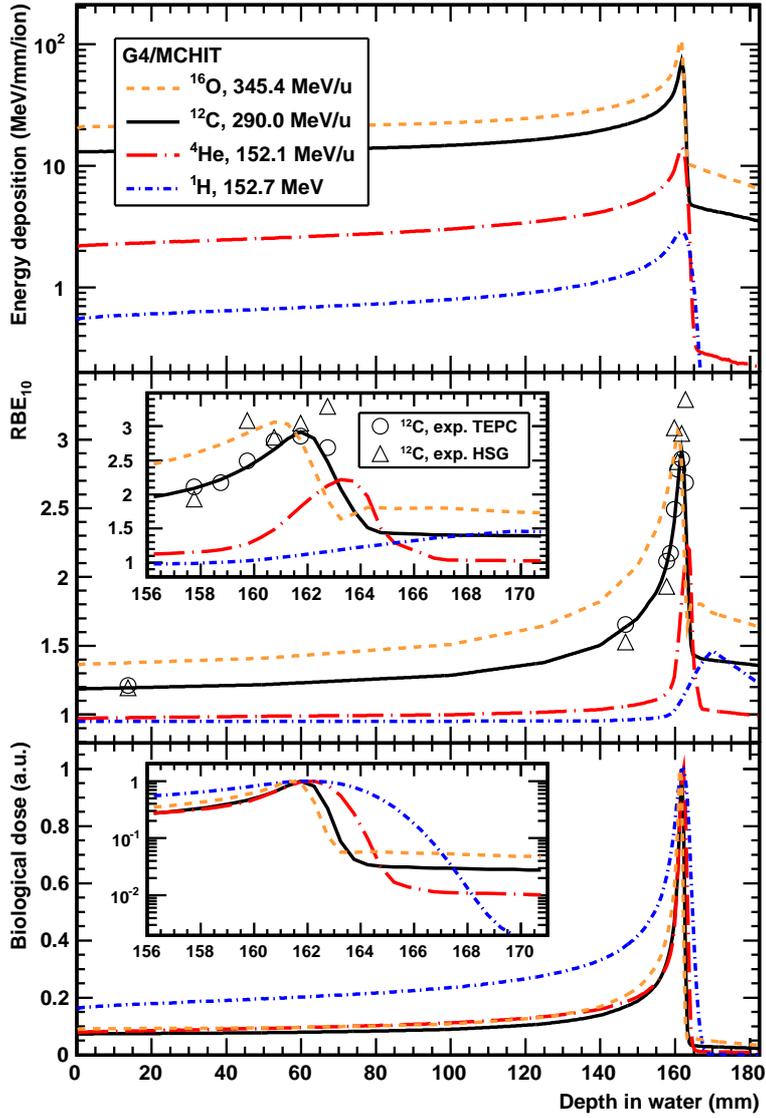}
\caption{Calculated energy deposition per beam particle (top panel), RBE$_{10}$ for HSG cells (middle panel) and biological dose (bottom panel) for $^{1}$H, $^{4}$He, $^{12}$C and $^{16}$O beams in water. The RBE$_{10}$ calculated from measured values of $y^*$~\cite{Kase2006} are shown by circles. The RBE$_{10}$ estimated from LQ fitting on survival curves of HSG cells~\cite{Kase2006} are presented by triangles. The biological dose distributions for all ions were rescaled to get the same value at the maximum.}
\label{fig:fig1}
\end{figure}

Simulation results for monoenergetic \unit{152.7}{\mega\electronvolt} $^{1}$H, \unit{152.1}{\mega\electronvolt\per\atomicmass} $^{4}$He, \unit{290}{\mega\electronvolt\per\atomicmass} $^{12}$C and \unit{345.4}{\mega\electronvolt\per\atomicmass} $^{16}$O beams are shown in
Figure~\ref{fig:fig1}. The beam energies were chosen to place the Bragg peaks of all four beams
at the depth of $\sim$\unit{161.8}{\milli\metre}.

As expected, the $^{12}$C and $^{16}$O beams deposit much higher energy at the Bragg peak compared to  $^{1}$H and $^{4}$He. However, higher energy deposition by $^{12}$C and $^{16}$O is also observed at the entrance and tail region. Therefore, 
no clear advantages of therapeutic $^{12}$C and $^{16}$O beams with respect to $^{1}$H and $^{4}$He beams 
regarding healthy tissues can be inferred 
exclusively from the analysis of the considered depth-dose distributions.

The RBE$_{10}$ profiles for the considered $^{1}$H, $^{4}$He, $^{12}$C and $^{16}$O beams are shown in the middle panel of Figure~\ref{fig:fig1} and they differ from each other. The RBE$_{10}$ for $^{12}$C beam estimated from the parameters of a LQ fitting on survival curves of HSG cells with $\beta$ fixed to 0.05~Gy$^{-2}$~\cite{Kase2006} is presented for comparison. The RBE$_{10}$ profile for $^{12}$C beam calculated with MCHIT coupled with the MK model agrees well with the profile which is also calculated with MK, but using the measured microdosimetry data on $y^*$~\cite{Kase2006}.
A prominent difference between RBE$_{10}$ profiles for $^{12}$C and $^{16}$O and the profile for $^{4}$He is seen in the insert
of Figure~\ref{fig:fig1}. The backward shift of the maximum of RBE$_{10}$ of $^{12}$C and $^{16}$O with respect to the 
position of the $^{4}$He maximum is due to the saturation effect. It is also found that $^{12}$C and $^{16}$O nuclei 
are characterized by higher RBE values for 10\% survival of HSG cells along the whole irradiated medium.
Their maximum values reach 2.9 and 3.1, respectively, close to the Bragg peak. At the same time the RBE$_{10}$ values for helium are relatively low at the entrance and tail regions and demonstrate a steep rise to 2.2 at the Bragg peak position. 
Before the Bragg peak the RBE$_{10}$ values for proton beam are slightly below 1. and increase to 1.5 well after the distal edge of 
the proton Bragg peak. This rise of RBE$_{10}$ for proton beam is explained by the presence of secondary nucleons produced
by beam protons in water and propagating beyond the Bragg peak.

Finally, the biological dose profiles for $^{1}$H, $^{4}$He, $^{12}$C and $^{16}$O (a.u.) are presented in the bottom panel of Figure~\ref{fig:fig1}. They were rescaled to get the same value at the maximum.
The profiles for $^{4}$He, $^{12}$C and $^{16}$O are very similar to each other. They are characterized by a more 
sharp rise and fall of the biological dose in the Bragg peak region compared to protons. After renormalization 
lower biological dose for $^{4}$He, $^{12}$C and $^{16}$O is predicted at 
the entrance with respect to protons. SOBP profiles of biological dose for all these projectiles 
are considered below.

\subsection{RBE distributions for $^{1}$H, $^{4}$He, $^{12}$C and $^{16}$O}

\begin{figure}[htb!]
\centering
\includegraphics[width=0.65\columnwidth]{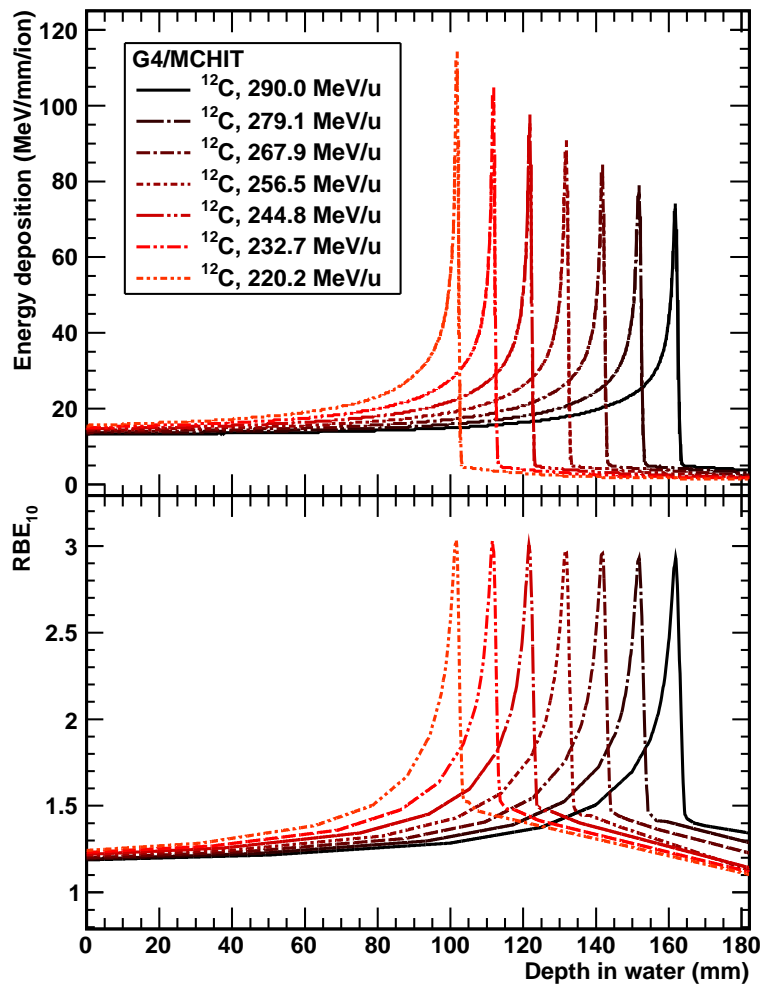}
\caption{Energy deposition profiles for $^{12}$C in water in \unit{10}{\milli\metre} steps  (top panel) and the 
corresponding RBE$_{10}$ profiles for HSG cells calculated with MCHIT and the MK model (bottom panel).}
\label{fig:fig2}
\end{figure}

As explained in Sec.~\ref{Sec:SOBPmethod}, a given biological SOBP dose distribution is composed from 
a set of depth-dose and depth-$y^*$ profiles calculated with MCHIT for monoenergetic beams.  
Such a library created in the present work contains pristine Bragg curves with a \unit{1}{\milli\metre} 
increment of the Bragg peak positions which are within 90--\unit{175}{\milli\metre} depth in water. 
Seven pristine Bragg peaks for $^{12}$C covering a \unit{60}{\milli\metre} domain in depth are shown 
in Figure~\ref{fig:fig2} to illustrate the content of this library.  
The microdosimetry variables which are also 
stored in the library as functions of depth are used to estimate RBE$_{10}$ profiles according to the MK model,
see Sec.~\ref{Sec:SOBPmethod}. 
The respective RBE$_{10}$ distributions are shown in Figure~\ref{fig:fig2} for the same beams. 
As seen in Figure~\ref{fig:fig2}, the height of the Bragg peak noticeably diminishes with depth, while the maximum 
RBE$_{10}$ remains almost constant ($\sim 3$) over the considered depth range. 

\begin{figure}[htb!]
\centering
\includegraphics[width=0.65\columnwidth]{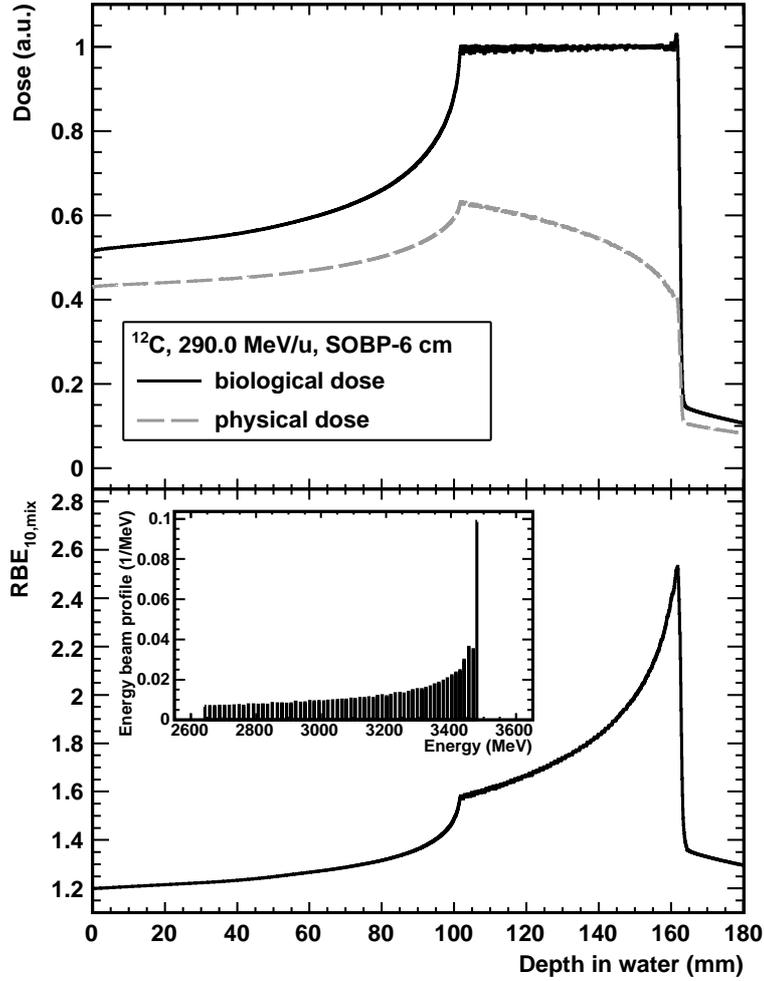}
\caption{A \unit{6}{\centi\metre}-wide SOBP biological dose distribution in water for 
\unit{290}{\mega\electronvolt\per\atomicmass} $^{12}$C ion beam and the respective physical dose (top panel). 
RBE$_{10,mix}$ profile for HSG cells (bottom panel) and the distribution of beam energy used to build
the SOBP profiles (insert).}
\label{fig:fig3}
\end{figure}

A \unit{6}{\centi\metre}-wide  SOBP profile of biological dose for \unit{290}{\mega\electronvolt\per\atomicmass} 
$^{12}$C beam, which was built according to the above-described procedure is shown in
the top panel of Figure~\ref{fig:fig3}. 
Here and in the following such profiles are normalized to 1 at the plateau in order to facilitate the comparison
of various ion species. The resulting distribution has a flat SOBP plateau with negligible 
fluctuations due to the presence
of individual Bragg peaks. In contrast, the corresponding SOBP distribution of the physical dose, which is also shown in  Figure~\ref{fig:fig3} is not flat, but rather decreases with depth. The respective RBE$_{10}$ 
amounts to $\sim 1.6$ at the proximal edge of the SOBP, while it is slightly above 2.5 at the distal edge, see the 
bottom panel of Figure~\ref{fig:fig3}.
The insert in the bottom panel of Figure~\ref{fig:fig3} demonstrates the calculated relative weights
for monoenergetic beams used to build the SOBP distribution of biological dose shown in Figure~\ref{fig:fig3}. 

\begin{figure}[htb!]
\centering
\includegraphics[width=0.65\columnwidth]{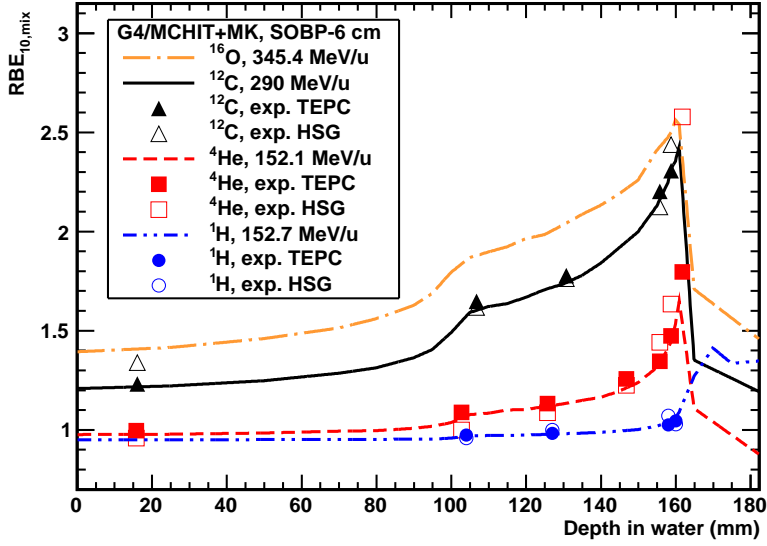}
\caption{RBE$_{10,mix}$ distributions calculated with MCHIT and the MK model for 
\unit{6}{\centi\metre} SOBP beams of $^{1}$H, $^{4}$He, $^{12}$C and $^{16}$O. The beam energy profiles for $^{1}$H, $^{4}$He and $^{16}$O were defined in such a way that they yield a \unit{6}{\centi\metre} SOBP at the same depth as for \unit{290}{\mega\electronvolt\per\atomicmass} $^{12}$C ion. The RBE$_{10,mix}$ values estimated with the MK model from measured $y^*$~\cite{Kase2006} are presented by full symbols as explained in the legend. The RBE$_{10,mix}$ values estimated from LQ fitting of survival curves of HSG cells are presented by open symbols.}
\label{fig:fig4}
\end{figure}

The \unit{6}{\centi\metre}-wide RBE$_{10,mix}$ profiles calculated for HSG cells for \unit{152.7}{\mega\electronvolt} $^{1}$H, \unit{152.1}{\mega\electronvolt\per\atomicmass} $^{4}$He, \unit{290}{\mega\electronvolt\per\atomicmass} $^{12}$C
and \unit{345.4}{\mega\electronvolt\per\atomicmass} $^{16}$O beams are presented in Figure~\ref{fig:fig4}.
They were calculated within the MK model basing on microdosimetry data generated by Monte Carlo simulations with MCHIT. 
The reliability of these profiles can be proved by comparing them with RBE$_{10,mix}$
calculated by two different approaches. In the first case, RBE$_{10,mix}$ is also calculated within the MK model, but on the basis of measured $y^*$ values~\cite{Kase2006}.
In the second approach, RBE$_{10,mix}$ is calculated using the parameters of LQ fitting of survival curves of HSG cells with $\beta$ fixed to 0.05~Gy$^{-2}$~\cite{Kase2006}.
The profiles based on MCHIT simulations agree very well with the RBE$_{10,mix}$ estimated on the basis of
the two set of experimental data, see Figure~\ref{fig:fig4}. In order to make such comparison, experimental values~\cite{Kase2006} 
corresponding to $^{1}$H and $^{4}$He were shifted in depth due to the difference of beam energies used in 
measurements and simulations. 

A good agreement with data for $^{1}$H, $^{4}$He and $^{12}$C suggests that this method can be also applied to $^{16}$O beam.
The distribution of RBE$_{10,mix}$ for $^{16}$O obtained on the basis of microdosimetry simulations
with MCHIT is shown in Figure~\ref{fig:fig4} for comparison. The shapes of RBE$_{10,mix}$ profiles for $^{4}$He, $^{12}$C and $^{16}$O are found to be similar with a rise of RBE for $^{12}$C and $^{16}$O at the proximal edge of the SOBP distribution.
A characteristic rise of RBE for $^{1}$H beyond \unit{165}{\milli\metre} depth at the distal region of the SOBP 
is found, similarly to the case of monoenergetic proton beam, Sec.~\ref{Sec:pristine}. 
It is found that RBE$_{10,mix}\sim 1$ at the entrance region of $^{1}$H and $^{4}$He beams.
The ratio between RBE$_{10,mix}$ values at the proximal (depth of $\sim$\unit{105}{\milli\metre}) and 
distal (depth of $\sim$\unit{165}{\milli\metre}) regions is larger for $^{16}$O compared to $^{12}$C. 
The RBE$_{10,mix}$ profile for $^{16}$O demonstrates the most pronounced tail with respect to other
projectiles. 

\subsection{SOBP distributions of biological dose}

\begin{figure}[htb!]
\centering
\includegraphics[width=0.65\columnwidth]{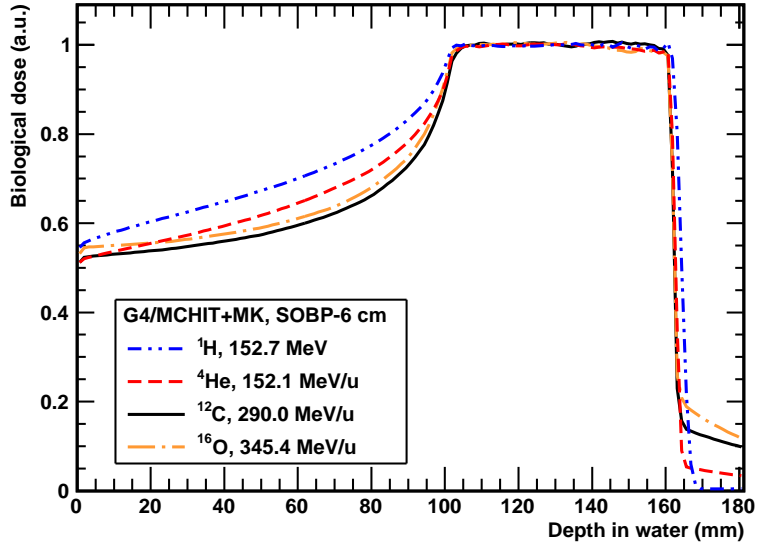}
\caption{SOBP biological dose profiles for $^{1}$H, $^{4}$He, $^{12}$C and $^{16}$O nuclei obtained 
as a product of physical dose calculated with MCHIT and RBE$_{10,mix}$ shown in Figure~\ref{fig:fig4}.
The distributions for all projectiles were rescaled to get the same value at the plateau.}
\label{fig:fig5}
\end{figure}

The SOBP distributions of biological dose $D_{bio}$ for $^{1}$H, $^{4}$He, $^{12}$C and $^{16}$O are shown 
in Figure~\ref{fig:fig5}. They are calculated for HSG cell as a product of physical dose calculated with MCHIT and RBE$_{10,mix}$ obtained within the MK model on the basis of microdosimetric modeling with MCHIT.
All four SOBP distributions presented in Figure~\ref{fig:fig5} are \unit{6}{\centi\metre} wide, and they 
are normalized to 1. at the plateau to facilitate the comparison of their shapes and ratios between the 
plateau and entry channel. Therefore, in the following the distributions of $D_{bio}$ are discussed in terms of 
dose relative to the plateau values. 

The distribution of $D_{bio}$ for $^{12}$C is characterized by the lowest values at the entrance channel. 
This helps to spare healthy tissues located in front of the target volume in treatments with $^{12}$C.
In contrast, the highest entrance dose is predicted for protons. However, the tail of the proton distribution 
beyond the distal edge of the plateau is negligible, while it is essential both for $^{12}$C and $^{16}$O. 
This indicates that the proton beam is the best option if very sensitive organs are located behind the tumor volume.
The choice of ion species for each specific treatment can, in principle, provide  
an optimal ratio between the doses in the entrance and tail regions. 

The distribution of $D_{bio}$ for $^{4}$He demonstrates a favorably small dose at the tail region, 
while its entrance value is higher compared to $^{12}$C. The biological dose delivered by $^{16}$O to normal tissue is slightly
increased both in front of the target volume and behind it compared to $^{12}$C. From the analysis of $D_{bio}$
distributions one can conclude that $^{12}$C 
is the best treatment option compared to $^{1}$H, $^{4}$He and $^{16}$O, unless only a 
very low $D_{bio}$ is acceptable in the tail region. In the latter case protons become the best treatment option.

\subsection{Distributions of cell survival fractions}

The central part of our study is devoted to the comparison of survival fractions of cells $S_{mix}$ 
calculated as a function of depth in the water phantom, which are estimated for tissues of different radiosensitivity. This makes possible to evaluate the respective therapeutic outcome for such tissues. 

\begin{figure}[htb!]
\begin{centering}
\includegraphics[width=0.75\columnwidth]{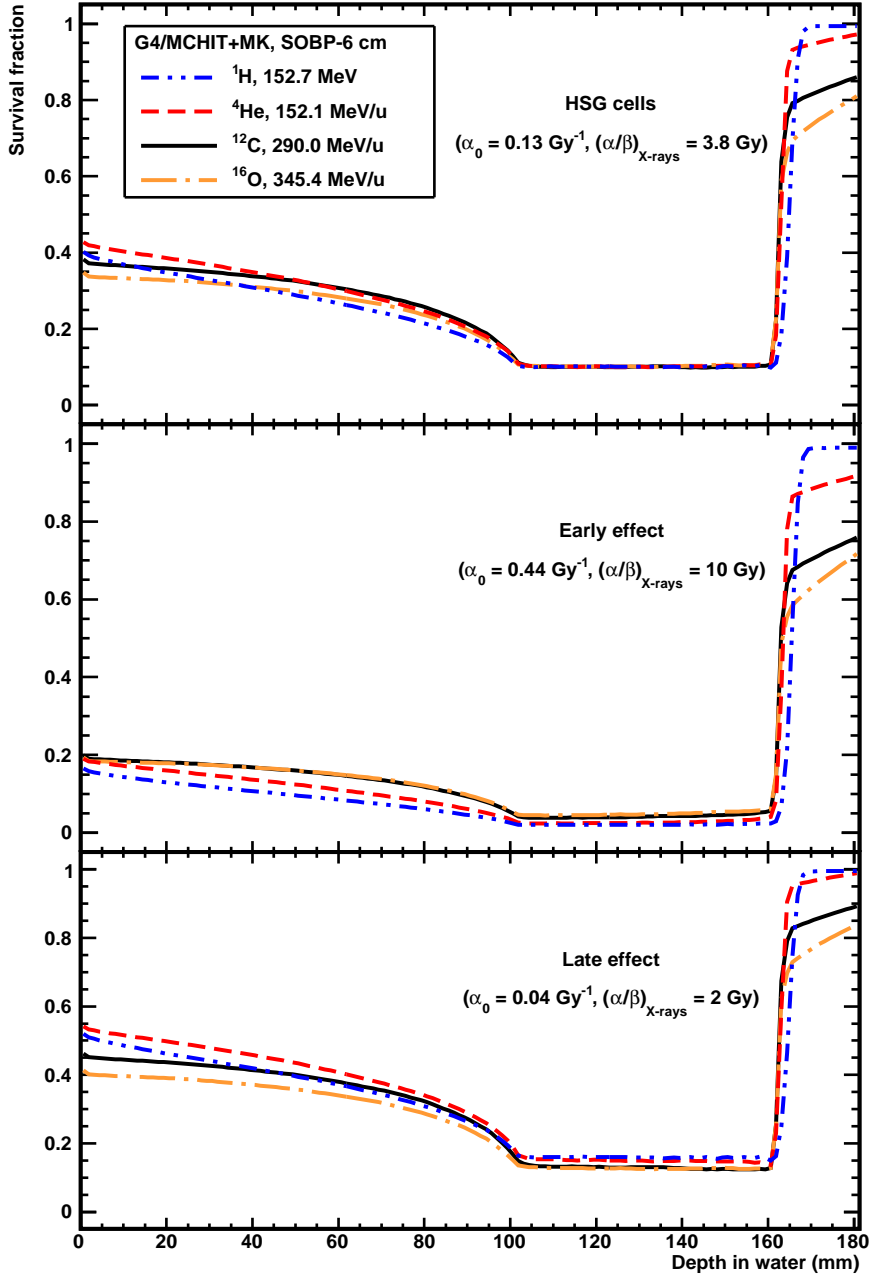}
\caption{Distributions of cell survival rates $S_{mix}$ after irradiation with $^{1}$H, $^{4}$He, $^{12}$C 
and $^{16}$O SOBP beams calculated for tissues with $(\alpha/\beta)_{X-rays}=3.8.$~Gy (HSG cells, top panel), 2~Gy (early responding tissue, middle panel) and 10~Gy (late responding tissue, bottom panel). The values of $\alpha_0$
parameter of the MK model are taken as 0.13~Gy$^{-1}$, 0.44~Gy$^{-1}$ and 0.04~Gy$^{-1}$, respectively.}
\label{fig:fig6}
\end{centering}
\end{figure}

The distributions of $S_{mix}$ calculated for cells (tissues) with $(\alpha/\beta)_{X-rays}=3.8.$~Gy, 2~Gy and 10~Gy 
after exposing them to $^{1}$H, $^{4}$He, $^{12}$C and $^{16}$O SOBP beams are shown in Figure~\ref{fig:fig6}.
Throughout this text the parameters for HSG cells were taken as $\alpha_0=0.13$~Gy$^{-1}$ and 
$(\alpha/\beta)_{X-rays}=3.8.$~Gy~\cite{Kase2006}. Hereafter the radiosensitivity of such tissues is considered
as normal. This serves as a natural reference point for comparison with two other tissues with their parameters taken following Kase {\em et al.}~\cite{Kase2011a}. The latter two cases correspond to
early responding tissue ($\alpha_0=0.44$~Gy$^{-1}$, $(\alpha/\beta)_{X-rays}=2$~Gy) very sensitive to radiation and
late responding tissue ($\alpha_0=0.04$~Gy$^{-1}$, $(\alpha/\beta)_{X-rays}=10$~Gy) which is radioresistant. 

The dose applied to HSG cells (normal radiosensitivity) leads to their 10\% survival at the target volume for all four beams.
The beams of $^{4}$He and $^{12}$C equally well spare tissue in the entrance channel, 
while the impact of $^{1}$H and $^{16}$O is stronger there. As one can expect, the main difference between 
light $^{1}$H, $^{4}$He and heavier $^{12}$C, $^{16}$O is revealed beyond the distal edge of SOBP profile
due to the tail of secondary fragments from $^{12}$C and $^{16}$O. One can also note that $^{16}$O is the 
worst option for HSG cells, while $^{4}$He is the best one.  
 
In the case of early and late responding tissues, see Figure~\ref{fig:fig6}, middle and bottom panels, 
the relation between the survival fractions estimated for $^{1}$H, $^{4}$He, $^{12}$C 
and $^{16}$O beyond the distal edge of the SOBP plateau are quite similar to the case of 
HSG cells. This is because of the fact that the dose in the tail region is defined by 
the presence of secondary fragments, which is essential for $^{12}$C and $^{16}$O beams.
In the case of early responding tissues less than 5\% of cells survive in the tumor volume, but their survival 
outside it is also unacceptably low (10--20\%). This means that in this case there are no advantages of 
charged particle therapy with respect to the intensity modulated radiation therapy (IMRT) with photons. 

In the case of late responding tissues more cells ($\sim 15$\%) survive in the tumor volume after the impact
of $^{1}$H, $^{4}$He, $^{12}$C and $^{16}$O SOBP beams, see Figure~\ref{fig:fig6}, bottom panel. The $^{12}$C and $^{16}$O 
SOBP beams are very effective in killing tumor cells, but the cell survival is also lower 
in the entrance channel ($\sim 40$\%) compared to $\sim 50$\% for $^{1}$H and $^{4}$He. 
By considering all three sensitivity cases one can conclude that the $^{4}$He beam is equally suitable for irradiation of
tissues with normal and low radiosensitivity as the $^{12}$C beam. Moreover, due to the reduced fragmentation 
of $^{4}$He, this option can be even better than $^{12}$C when sparing tissues after the tumor volume is crucial. 
At the same time the $^{16}$O beam has no clear advantages compared to $^{12}$C. Due to higher ionization in the
entrance channel and enhanced fragmentation more cells are killed by $^{16}$O beam outside the tumor volume 
compared to $^{12}$C beam. However, one can consider $^{16}$O is a good option for highly resistant tumors, as it 
effectively kills cells in the tumor volume.

\section{Conclusions}

In this work we presented our approach based on Monte Carlo simulations of microdosimetric spectra of 
monoenergetic beams of $^{1}$H, $^{4}$He, $^{12}$C and  $^{16}$O in water. It provides $y^*$ values 
as input to the modified MK model~\cite{Kase2011} for calculating RBE$_{10,mix}$ for HSG cells for 
SOBP distributions composed from  
monoenergetic beams of these projectiles. This method gives RBE$_{10,mix}$ for $^{1}$H, $^{4}$He, $^{12}$C 
which are in full agreement with RBE$_{10,mix}$ also calculated within the MK model, but from measured $y^*$~\cite{Kase2006}
and RBE$_{10,mix}$ calculated from the parameters of LQ fitting of survival curves of HSG cells~\cite{Kase2006}.
This makes us confident in extending our approach to $^{16}$O beams for which the respective data are not 
available. Our approach provides well-adjusted biological dose distributions for $^{1}$H, $^{4}$He, $^{12}$C and $^{16}$O with
a very flat SOBP plateau. Thus basic properties of mixed radiation fields in treatments with these projectiles are
emulated.

It is found that the shapes of RBE$_{10,mix}$ profiles for $^{4}$He, $^{12}$C and $^{16}$O are similar to each other,
while the RBE$_{10,mix}$ for protons is almost constant ($\sim 1.$) over the whole depth  in water,  
excluding enhanced RBE$_{10,mix}$ ($\sim 1.2$) after the distal edge of the SOBP plateau.
Considerably lower RBE$_{10,mix}$ values are estimated in the entrance and tail region for $^{4}$He compared 
to  $^{12}$C and $^{16}$O. In the target volume the highest RBE$_{10,mix}$ values of 1.5--2.5 are calculated for 
$^{12}$C and $^{16}$O. 

In order to reduce side effects of ion therapy such as radionecrosis~\cite{Kase2011a} the damage to 
surrounding healthy tissues should be reduced as much as possible. With the help of our MCHIT model 
connected with the modified MK model the severity of this damage is evaluated 
by calculating the cell survival fractions in healthy tissues for 
several kinds of therapeutic beams ($^{1}$H, $^{4}$He, $^{12}$C, $^{16}$O). 
We considered the cases of normal (HSG cells), high and low 
radiosensitivity of tissues in the tumor volume and around it. 
The consideration of the impact of $^{1}$H, $^{4}$He, $^{12}$C and $^{16}$O SOBP beams in these three
cases led us to the following conclusions:
\begin{itemize}
\item In the case of early responding tissues all four charged particle beams induce 
severe damage not only to the target volume, but also around it. Since in this case the region of high damage
is not conformal to the target volume, the treatment with charged particles loses its advantages with 
respect to treatment with photons.
\item In the case of tissues with normal radiosensitivity (HSG cells) $^{4}$He and $^{12}$C beams spare tissue 
in the entrance channel better than $^{1}$H and $^{16}$O ones.
\item $^{4}$He and $^{12}$C nuclei are equally suitable for irradiation of
tissues with normal and low radiosensitivity. The cell survival fractions calculated, respectively, for the entrance
channel and target volume are similar for $^{4}$He and $^{12}$C.  
\item However, as soon as it is important to spare healthy tissues after the distal edge of the SOBP plateau,
$^{4}$H can be recommended due to the reduced nuclear fragmentation of these projectiles.
\item No definitive advantages of $^{16}$O with respect to $^{12}$C were found, with the except of an enhanced impact
of these heavier projectiles on radioresistant tumors. 
\end{itemize}

In a recent work~\cite{Remmes2012} the authors studied the possibility to spare healthy tissues by properly
selecting ion species for therapy. In addition to $^1$H, $^4$He, $^{12}$C they considered lithium, beryllium, boron and 
neon ions and calculated the dose to normal tissue delivered by these beams.  
However, the option of $^{16}$O was not considered, while the Heidelberger Ionenstrahl-Therapiezentrum 
(HIT) in Heildeberg, Germany, provides $^{16}$O beams of therapeutic energies~\cite{Haberer2004,Combs2010a} in addition
to $^1$H, $^4$He, $^{12}$C.  Treatments at this facility are performed presently only with protons and carbon ions, but
$^{16}$O can be also used following respective pre-clinical studies.  
In this sense our study complements the results of Remmes {\em et al.}~\cite{Remmes2012} by 
considering $^{16}$O beams, and also by comparing cell survival profiles in $^{1}$H, $^{4}$He, $^{12}$C and 
$^{16}$O treatments in addition to biological dose profiles. 

As suggested~\cite{Bassler2010}, multi modal irradiations with various nuclei can be used for LET-painting.
In this method high-LET radiation is used to boost the LET in a hypoxic sub-volume of the target (hypoxic compartments of the tumor). At the same time low-LET radiation is applied to the complementary target volume. Such combination may increase tumor control and reduce 
side effects. This means that a thorough evaluation of the physical properties and biological effectiveness of 
different beams is necessary before they can be applied in real treatments. Our approach can be also used for estimating RBE and cell survival fractions for $^{7}$Li, $^{8}$Be, $^{10}$B and $^{14}$N nuclei prior to planning radiobiological experiments with these beams.

\section*{Acknowledgements}
This work was partially supported by HIC for FAIR within the Hessian LOEWE-Initiative.
Our calculations were performed at the Center for Scientific Computing (CSC) of the
Goethe University Frankfurt. We are grateful to the CSC staff for support.

\bibliographystyle{unsrt}

\end{document}